\documentclass[conference,letterpaper,final,draft]{IEEEtran}

\usepackage[utf8]{inputenc}
\usepackage{color}
\usepackage{textcomp}
\usepackage{url}
\usepackage{amsmath}
\usepackage{graphicx}
\usepackage{balance}
\usepackage{booktabs}
\usepackage{cite}
\usepackage[font=small]{caption}

\IEEEoverridecommandlockouts
\makeatletter
\let\@period=\@empty
\makeatother

\usepackage{array}
\newcolumntype{L}[1]{>{\raggedright\let\newline\\\arraybackslash\hspace{0pt}}m{#1}}
\newcolumntype{C}[1]{>{\centering\let\newline\\\arraybackslash\hspace{0pt}}m{#1}}
\newcolumntype{R}[1]{>{\raggedleft\let\newline\\\arraybackslash\hspace{0pt}}m{#1}}
\newcolumntype{P}[1]{>{\centering\arraybackslash}p{#1}}

\let\oldcaption = \caption
\renewcommand{\caption}[1]{\oldcaption{#1}\vspace{-5mm}}

\makeatletter
\def\section{\@startsection{section}{1}{\z@}{1.5ex plus 0ex minus 0ex}%
{1sp}{\normalfont\normalsize\centering\scshape}}%
\def\subsection{\@startsection{subsection}{2}{\z@}{1.5ex plus 0ex minus 0ex}%
{1sp}{\normalfont\normalsize\itshape}}%
\def\subsubsection{\@startsection{subsubsection}{3}{\parindent}{0ex plus 0ex minus 0ex}%
{0ex}{\normalfont\normalsize\itshape}}%
\def\paragraph{\@startsection{paragraph}{4}{2\parindent}{0ex plus 0ex minus 0ex}%
{0ex}{\normalfont\normalsize\itshape}}%

\newcounter{protocol}[subsection]

\def\protocol{\@startsection{protocol}{3}%
{\z@}{1pt}{1pt}
{\normalfont\normalsize\itshape\bfseries}}
\makeatother

\begin{document}
\title{LocaRDS: A Localization Reference Data Set}

\author{
\IEEEauthorblockN{Matthias Schäfer\IEEEauthorrefmark{1},
Martin Strohmeier\IEEEauthorrefmark{2}\IEEEauthorrefmark{4},
Mauro Leonardi\IEEEauthorrefmark{3},
Vincent Lenders\IEEEauthorrefmark{4}}\\[-0.3em]
\IEEEauthorblockA{
\begin{tabular}{C{0.23\textwidth}C{0.23\textwidth}C{0.23\textwidth}C{0.23\textwidth}}
 \IEEEauthorrefmark{1}TU Kaiserslautern, Germany&%
 \IEEEauthorrefmark{2}University of Oxford, UK & %
 \IEEEauthorrefmark{3}University of Rome, Tor Vergata, Italy & %
 \IEEEauthorrefmark{4}armasuisse S+T,  Switzerland
\end{tabular}}
}

\maketitle

\begin{abstract}

The use of wireless signals for purposes of localization enables a host of applications relating to the determination and verification of the positions of network participants, ranging from radar to satellite navigation. Consequently, it has been a longstanding interest of theoretical and practical research in mobile networks and many solutions have been proposed in the scientific literature. However, it is hard to assess the performance of these in the real world and, more severely, to compare their advantages and disadvantages in a controlled scientific manner.

With this work, we attempt to improve the current state of the art in localization research and put it on a solid scientific grounding for the future. Concretely, we develop LocaRDS, an open reference dataset of real-world crowdsourced flight data featuring more than 222 million measurements from over 50 million transmissions recorded by 323 sensors. We show how LocaRDS can be used to test, analyze and directly compare different localization techniques and further demonstrate its effectiveness by examining the open question of the aircraft localization problem in crowdsourced sensor networks. Finally, we provide a working reference implementation for the aircraft localization problem and a discussion of possible metrics for use with LocaRDS.

\end{abstract}

\section{Introduction}

In mobile wireless networks, three related problems have received considerable attention in the scientific literature and industry applications over the past decades: positioning (or localization), self-positioning, and location verification. In \textit{positioning}, the wireless network aims to determine the concrete location of a remote and typically mobile node, whilst in \textit{self-positioning}, the remote node seeks to find out its own (absolute or relative) position. A common pattern in systems where self-positioning is used to establish location awareness among nodes is that location information is exchanged across the network. In scenarios where location is used in safety- or security-critical services, the need for \textit{location verification} arises. Here, a set of trusted nodes aims to verify the correctness of locations reported by untrusted nodes.

Widely-deployed examples of positioning or verification schemes are radar systems, which are utilized by civil or military traffic management to find the positions of aircraft or ships. In turn, Global Navigation Satellite Systems are used for self-positioning by billions of devices and users around the world. These technologies also provide location awareness to a diverse set of applications such as traffic control, environmental monitoring, or emergency services. Fueled by the global and free availability of GPS, self-positioning has become particularly popular in the transportation domain due to its excellent scalability and coverage. This development is accompanied with the need for location verification, especially in safety critical domains such as transportation \cite{van2018veremi,strohmeier2018k}.

One common approach to solving all three problems is to exchange signals between a node at an unknown or untrusted position and multiple nodes at known positions. Based on measurable physical signal effects such as propagation delay or path attenuation, the target node's position is then estimated (or verified) by inserting the measurements into a model and solving for the unknown location.

\begin{figure}[t]
    \centering
    \includegraphics[width=0.75\linewidth]{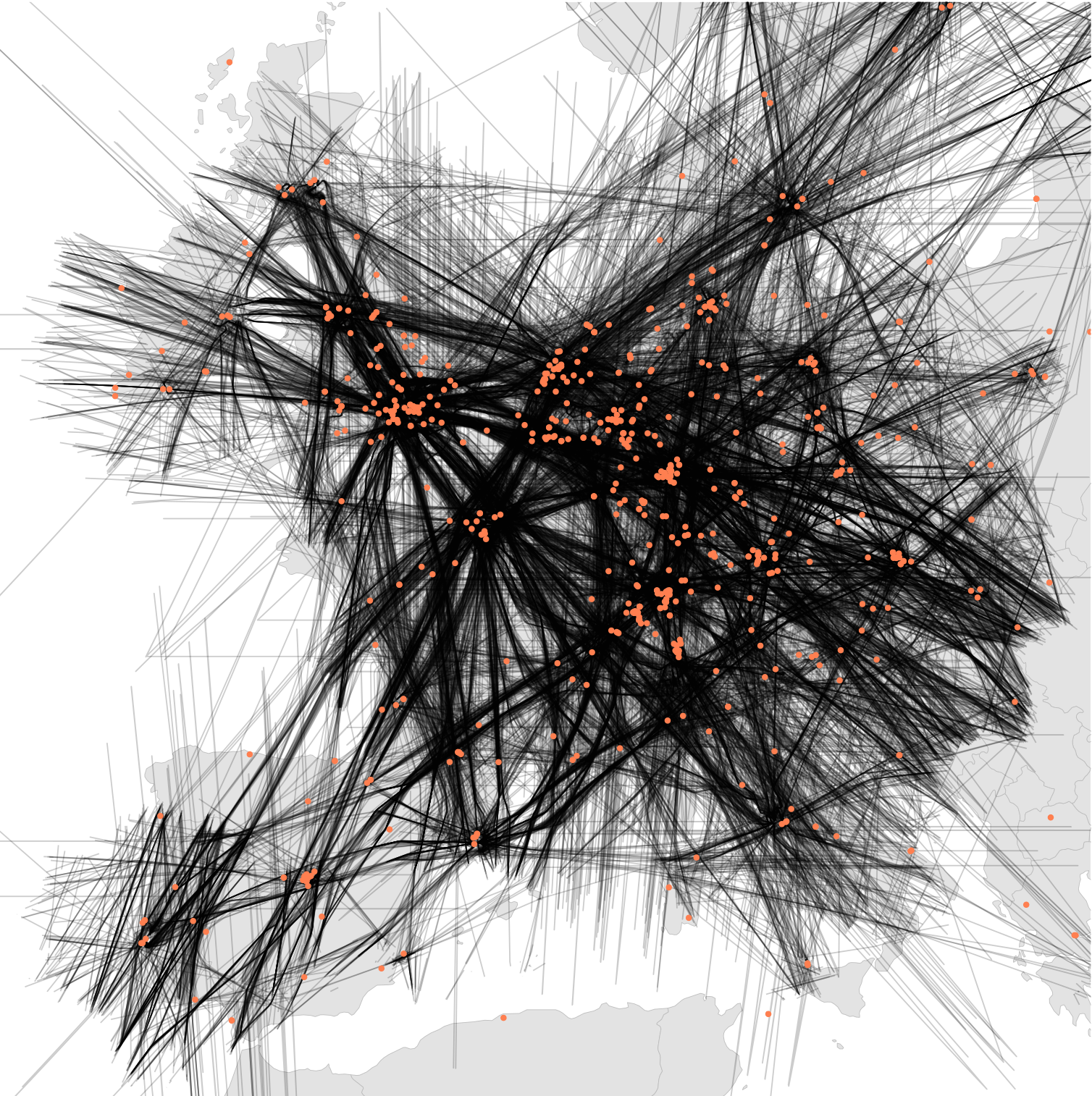}
    \caption{The LocaRDS dataset with 50,865,291 aircraft positions (black lines) and 323 sensor positions (orange dots). In addition to this geographical information, the dataset contains signal strength and time of arrival measurements for each position reported by an aircraft.}
    \vspace{-5pt}
    \label{fig:locards_locations}
    
\end{figure}

Generally, localization solutions discussed in the literature are evaluated either using simulations or unique and difficult to replicate lab setups and scenarios. Consequently, it is hard to assess their performance in the real world and, more severely, to maintain the core scientific principles of reproducibility and comparability. Live competitions have emerged as one potential solution to this problem for indoor localization, pitting a number of approaches against each other in a realistic setting \cite{lymberopoulos2015realistic,van2013evarilos}. However, besides inherent difficulties in replicating such one-time events for non-participating approaches, they are very resource-intensive for organisers and participants, leading to the exclusion of many potential solutions. 

We argue that there are two main reasons for the lack of repeatable and realistic evaluations in outdoor localization research. First, real-world measurements can be expensive in domains such as aviation or shipping as multiple high-quality receivers are needed, which cost each up to 1000 USD or more. Second, researchers often lack access to a good variety of  multiple sites over a large area as organizations with such infrastructure are generally not allowed to share their data.

Open and crowdsourced initiatives such as the OpenSky Network \cite{schafer2014bringing} exist to address these issues and have done much to improve the availability of wireless data from their domain. However, their focus is generally on the data contents. Access to and usage of physical layer information requires deep system and domain knowledge for the necessary complex pre-processing and data cleansing. Thus, their utility is currently limited for research into localization algorithms, exacerbated by the large size of these initiatives, which does not guarantee that a chosen sample is representative and comparable. 

Altogether, this leads to a complete fragmentation of localization research, where authors have to build their own test sets from real or simulated data in order to compare their novel methods. In this paper, we solve this problem by presenting LocaRDS, a prepared and proven reference dataset, which fulfils the scientific requirements of many localization methods. By using the established research network OpenSky, we create an open dataset of real-world crowdsourced flight data, containing crowdsourced physical layer information of more than 50 million messages gathered from 323 sensors distributed over a large and diverse geographical area in Europe (see Fig. \ref{fig:locards_locations} for a graphical illustration).  The dataset is suited for any research in the areas of localization, location verification, positioning or other fields where time difference of arrival and signal strength measurements are needed.

Our larger mission with this work is to improve the state of the art in long-distance localization research and put it on a solid scientific grounding for the future. We hope by using a comparable and standardized reference source, it becomes clear which are the best solutions to these problems under realistic conditions. Related to this goal, we discuss the requirements and the metrics, which need to be implemented for such a dataset, in order to enable a fair and useful comparison.

\subsection*{Contributions}

\begin{itemize}

    \item We develop LocaRDS, a novel reference dataset for effective scientific comparability in localization research based on crowdsourced real-world air traffic data.
    
    \item We apply LocaRDS in a case study, implementing a solution for the aircraft localization problem (ALP) in heterogeneous, crowdsourced receiver systems. 
        
    \item We discuss evaluation metrics in order to establish a new benchmark system for improved solutions of the ALP.
\end{itemize}

\emph{The full dataset is available at \url{https://doi.org/10.5281/zenodo.4298998} \cite{zenodoLocaRDS} under the CC BY-SA license. The authors are members of the OpenSky Network association.}

\section{Application Areas}\label{sec:background}

LocaRDS targets research where the following system and communication characteristics are common:
\begin{itemize}
    \item \textbf{Location awareness:} Location information is needed and/or shared between network nodes. This often implies that (some) nodes are \emph{mobile} since there is less need for constant location awareness in static networks.
    \item \textbf{Wireless communications:} Nodes use point-to-point radio links to establish and maintain location awareness, specifically in line-of-sight environments. Such wireless communications are typical, e.g., in mobile ad hoc networks.
    \item \textbf{Passive infrastructure:} Radio signals are received by a set of passive sensors. Moreover, a single signal should be received by multiple sensors.
    \item \textbf{Signal metadata:} Each receiver measures signal strength and/or time of arrival (ToA). These metadata can (but do not have to) be synchronized or otherwise calibrated.
\end{itemize}

LocaRDS is specifically targeted at (yet, not limited to) three research domains where these characteristics are common: wireless positioning, self-positioning, and location verification. While all three methods can in principle be conducted with a wide range of physical measurements, such as Doppler shift or angle of arrival, we focus on the most popular ones: received signal strength (RSS) and time difference of arrival (TDoA). 

\subsection{Positioning}
As wireless positioning is a widely applicable and popular research area, providing a complete overview here is beyond the scope of this work. Differences in physical characteristics, algorithmic techniques and application scenarios are is covered in detail in several books and surveys.

As an introduction to the topic, Mao and Fidan \cite{mao2009localization} give an extensive overview of the classical strategies and algorithms used in this area. The authors in \cite{guvenc2009survey} further systematize 29 ToA-based techniques, in particular in non-line-of-sight environments, while \cite{lymberopoulos2015realistic} reports from an indoor competition comparing 22 different teams. 
However, of particular interest for us are positioning works such as \cite{strohmeier2018k,WAM}, which consider passive outdoor line-of-sight environments as they are found for example in the data underlying LocaRDS. 

\subsection{Self-Positioning}

The goal of positioning (or localization) is generally to establish awareness of locations of \emph{remote} nodes. Another, closely related but different problem is enabling a node to determine its own location. We explicitly distinguish the two problems by referring to the latter as self-positioning. Self-positioning approaches usually employ one or more moving beacon nodes which constantly broadcast signals. These signals are used by passive nodes to infer their own locations. Global navigation satellite systems such as GPS are a prominent real-world example for this system design.

Academic works on self-positioning (e.g., \cite{Luo06,Eichelberger17}) are mostly focused on indoor scenarios, likely due to global GPS availability outdoors. By relying on aircraft signals, LocaRDS is perfectly aligned with this system design. In fact, the evaluation in \cite{Eichelberger17} uses OpenSky measurements similar to LocaRDS, although at a much smaller scale and not public.

\subsection{Secure Location Verification}

In location verification, a set of nodes aims at verifying the correctness of location information provided by untrusted nodes \cite{Sastry03}. Approaches follow a general pattern where location-dependent and tamper-proof physical signal propagation characteristics are measured and compared to a claimed location. If the measurements do not comply with the claimed distance, location, track, or region, the information will be rejected by the system. Existing location verification methods can be broadly classified into angle of arrival, time of flight (or distance bounding), and time difference of arrival approaches.

While LocaRDS can support all three classes, e.g., by serving as input for simulations with realistic node movement and communication behavior, the class that can arguably benefit the most are TDoA-based approaches \cite{Capkun04, Capkun08, Strohmeier15, Schaefer15, Baker16, Schaefer19}. They rely on timestamps for the arrivals of a signal at different locations, something LocaRDS provides. Similarly, research in data validation such as \cite{Cycon17Strohmeier} and \cite{schafer2018opensky} could benefit from LocaRDS, especially those with a focus on crowdsourced data.

\subsection{Receiver Synchronization}
The calibration and synchronization of the receiving sensors itself is also an intermittent subject of research as it is a key requirement across all three discussed research areas. Many of the solutions in these domains require very tight time synchronization, in particular those based on TDoA measurements. This is costly even in controlled industrial deployments but impossible to achieve consistently with the variety of modern crowdsourced sensors used by enthusiasts to feed OpenSky and similar networks. In OpenSky, about 80\% of all feeders use cheap RTL-SDR and Raspberry Pi setups, which do not support GPS synchronization and cost broadly in the range of \$50-100. Conversely, fewer than 20\% use setups which are about one order of magnitude more expensive and offer relatively tight synchronization via GPS. Moreover, the accuracy of these setups is dependent not only on the availability and quality of time synchronization but also bounded by the resolution of the internal clocks and the sampling process. The latter can vary significantly from 2.4~MHz for an RTL-SDR dongle vs. about 60~MHz for a more expensive Radarcape) and thus influences the level of noise time-based localization solutions have to deal with.

Consequently, any positioning method working in such an uncontrolled, mixed-sensor environment must be able to process timestamp data of varying accuracy and provide a good robustness against outliers. For example, the authors in \cite{strohmeier2018k} develop a solution based on the k Nearest Neighbor algorithm and show that it is less susceptible to noisy timestamps and imperfect sensor network geometry than traditional MLAT approaches. Due to the dilution of precision effect in these algorithms, the issue of inaccurate timestamps becomes more problematic in crowdsourced deployments or, in general, where the relative positions of receivers and localization targets are non-optimized. For example, in  \cite{mantilla1} this problem was solved for  MLAT implementation outside the airport, using a regularization method for the ill-posed problem.

\section{Dataset Requirements}\label{sec:requirements}
After summarizing the design aspects of positioning, self-position\-ing, and location verification, we derive the requirements for the dataset covering both the requirements with regards to the underlying system and the researcher requirements. We cover the overarching motivation of scientific comparability in detail in the following section.

\subsection{System Requirements}
For a consistent, useful dataset, these requirements should be fulfilled by the underlying collection and processing system.

\begin{itemize}
    \item \textbf{Real-world data:} Synthetically created data is not able to capture all complexities found in the variety of receiver setups and their respective environments. It is difficult to accurately model all factors influencing even a single controlled wireless receiver, let alone a host of heterogeneous and uncontrolled ones. Hence, only real-world data can provide the basis for our dataset as it contains measurements which include all uncountable factors that influence communications and measurements in a realistic scenario.
    
    \item \textbf{Large-scale deployment:} To be able to conduct effective localization with sufficiently overlapping node coverage in various geographic and geometric scenarios, a large-scale deployment with hundreds of nodes with redundant coverage in most areas is crucial.
    
    \item \textbf{Variety in hardware and software:} To show the generality and wide applicability of the solution approaches, there should be different sensor types with different capabilities (e.g., time synchronization) providing the underlying data. Similarly, different software features and typical behaviour should be reflected in the dataset.

\end{itemize}

\subsection{Researcher Requirements}
In the following, we consider the needs of the research users and what is necessary to guarantee the widest possible accessibility of LocaRDS, without which the goal of scientific comparability would fall short.
\begin{itemize}
    \item \textbf{Ground truth available:} To judge the accuracy of an approach, there must be ground truth about both the origin and the destination of a signal.

\item \textbf{Documentation:} To understand the dataset and use it effectively in research, the underlying collection system must be described and all possible data and system artifacts should be known and documented.
    
\item \textbf{Pre-processing:} The data needs to be preprocessed to remove unnecessary information and prepared in a form to make it readily usable for a wide range of applications; the standard CSV format being the natural choice.

\item \textbf{Variable scenarios:} Localization research comes in many different forms, using a heterogeneous mixture of physical layer data and message content. These scenarios should be covered as well as possible, providing different parameters (e.g., states of synchronization, location information, mobile nodes, stationary nodes, different accuracy levels of ground truth, different geometries, different numbers of receivers). 
    
\item \textbf{Large number of measurements:} Modern machine learning solutions, in particular those using deep learning (e.g., \cite{niitsoo2018convolutional}) require a large numbers of raw measurements to use in training to be effective. This means providing millions of measurements to work with in any particular setting.

\item \textbf{Quality indicators:} To make selection of subsets for specific scenarios easier, there needs to be an estimate about the quality of any particular measurement. This includes, for example, the quality of synchronization, clocks or timestamps.

\item \textbf{Privacy:} To comply with ethical requirements, the privacy of all participating nodes must be respected.
\end{itemize}

\subsection{Scientific Comparability}

We follow in the footsteps of recent efforts in the networking, transportation and security research communities (for example \cite{van2018veremi}), which aim for a better reproducibility of scientific experiments, after the rise of what has been dubbed a ``reproducibility crisis'' in many scientific fields \cite{baker2016there}.

Existing work relating to the positioning research has generally relied on individually-designed studies, either based on simulations or with data collected using the methods available to researchers and their collaborators. Typically, both code and dataset(s) are not available, often due to intellectual property rights or other constraints imposed for example by industry collaborators in sensitive fields such as defense or aviation. While this approach means that the solution to positioning problems can be tailored specifically to the scenario as desired by collaborators or customized to the environment of the researchers, it makes it impossible in practice to compare different approaches to each other. Previously, the Evarilos project has attempted to improve this issue in indoor localization by providing comparative evaluation scenarios in healthcare and underground mining \cite{van2013evarilos}. However, the benchmark suite and data has become unavailable since.

Hence, the purpose of our dataset is to provide a common baseline to compare large-scale localization solutions in the real world. This has several positive effects, from reducing the time and effort for researchers to perform high-quality studies to better comparability for users interested in the advantages and disadvantages of the available solutions. While the use of LocaRDS cannot replace the in-depth analysis of a novel localization approach, we believe that the existing situation is a significant hindrance to the advancement of the field. Currently, a researcher seeking to compare their approach with any other needs to conduct time- and resource-intensive re-implementations, which are often difficult and error-prone due to the lack of low-level implementation information.

LocaRDS' aim is to provide the first step towards a comprehensive evaluation methodology for this field and we hope other subdomains in the wireless localization field follow this example, thereby creating a larger repository of verified data.

\section{LocaRDS: The Localization Reference Dataset}\label{sec:dataset}

To meet the requirement of documentation, we describe how LocaRDS was collected and prepared. This knowledge is key to understanding which processing and system artefacts are to be expected in the data. We will then provide an overview of the structure and contents of LocaRDS.

\subsection{Collection}
\label{sec:collection}

The raw data used to generate LocaRDS was recorded by the OpenSky Network (OSN) \cite{schafer2014bringing}. The OSN is a network of more than 2500 crowdsourced sensors which collects air traffic control data at a large scale and it provides these data to researchers for free. The network records the payloads of all 1090~MHz secondary surveillance radar downlink transmissions of aircraft along with the \emph{timestamps} and \emph{signal strength indicators} provided by each sensor on signal reception.

Part of this dataset are the exact aircraft locations that are broadcast twice per second by transponders using the Automatic Dependent Surveillance--Broadcast (ADS-B) technology. With respect to the requirements discussed in Section \ref{sec:requirements}, we are only interested in this position reporting subset of ADS-B as they provide the GNSS-derived locations of the transmitters which serve as the necessary ground truth. In addition to this measurement data, the OSN provides a list of all sensors, including their location and the device type.

Together, these two datasets (measurements and sensor information) constitute a well-suited basis for LocaRDS. In particular, we highlight several properties relevant to localization, self-positioning, and location verification problems:
\begin{itemize}
    \item \textbf{Known locations:} The locations of all nodes (transmitters and receivers) are known. This provides the ground truth and reference locations needed by many algorithms.
    
    \item \textbf{Coverage redundancy:} Each transmission is received by multiple receivers. Since this is a key requirement for most localization algorithms, we limit LocaRDS to data recorded in Central Europe where OpenSky's redundancy is highest and its coverage is nearly complete.
    
    \item \textbf{Diversity:} There are thousands of different transmitters constantly broadcasting their locations while hundreds of different receivers are recording these signals. This provides a rich set of measurements with varying accuracy and geometry, which in turn allows researchers to test the influence of different factors on the performance of their algorithms.
    
    \item \textbf{Mobility:} The transmitting nodes (i.e., the aircraft) are moving through the network at variable and typically very high speeds of 800 km/h and more, creating a highly dynamic network topology.
    
    \item \textbf{Crowdsourced:} The data comes from a crowdsourced system of receivers, integrating all the challenges and difficulties found in such an organically grown, non-controlled set of receivers. Collecting data from a synchronized and controlled deployment would be less flexible and less widely applicable. Conversely, due to the high sensor density and high level of redundancy in the OSN, selected subsets of this data can emulate a controlled deployment.
\end{itemize}

\subsection{Preparation}
\label{sec:preparation}

To make the data accessible and meet the requirements, complex pre-processing is needed to abstract from most system aspects, reduce the data volume, and to eliminate the need to understand all system aspects in order to use the data. Moreover, the information quality needs to be assessed and indicated, allowing researchers to choose subsets that match their own requirements. Therefore, we performed the following processing steps to prepare the unstructured OSN data and create a well-defined dataset for scientific comparisons:

\subsubsection{Decoding} Decoding ADS-B correctly is a complex task. Although libraries and tutorials such as \cite{Sun19} exist, it remains a tedious task that requires a deep understanding of the underlying link layer technology Mode~S. Moreover, the sheer volume of data collected by OpenSky (about 120~GB of raw data per hour) makes this process  challenging and resource-intensive. Therefore, we relieve researchers from this burden by providing readily decoded location information in WGS84 coordinates, altitude information in meters, and a unique aircraft identifier as a simple integer.

\subsubsection{Continuous timestamps} Timestamps are provided in different resolutions and units, depending on the receiving sensor type. Moreover, some sensors only provide rolling counters, counting from 0 to some power of two at a certain frequency. To abstract from this, we implemented a counter overflow detection mechanism and mapped all timestamps to a continuous timestamp with a common unit (nanoseconds). The complexity of this preparation step comes from the fact that the raw data coming from a sensor can have an unknown and varying delay, making the estimation of overflows difficult.

\subsubsection{Deduplication} OpenSky's raw data is merely a long list of single measurements by single sensors. However, as most localization algorithms rely on signals being received by multiple receivers, we grouped multiple receptions belonging to the same transmission based on their continuous timestamp and signal payload. This process is called deduplication. Note that although most position reports are unique, a small number of falsely grouped measurements remains as noise in the data.

\subsubsection{Filtering} To make the dataset more manageable without losing temporal or topological properties, we reduced the input dataset to ADS-B position reports received in Europe, where OpenSky's coverage and redundancy is by far best. Moreover, we discarded transmissions received by a only one sensor since they have little value for the purposes of LocaRDS and omitting them reduces the size of the dataset significantly.

\subsubsection{Quality assessment} Crowdsourcing creates several issues regarding the quality and integrity of location and timing information of certain aircraft and sensors \cite{schafer2018opensky}. To allow researchers to ignore these effects while still preserving them as a potential subject of research, we implemented integrity checks from \cite{schafer2018opensky} to verify and judge the data correctness and added respective indicators to LocaRDS. In principle, our data validation first detects GPS-synchronized sensors with accurate location information based on their timestamps. These GPS-synchronized timestamps are then used to verify the location information provided by the different aircraft.

\subsection{Available Information}

The following information are available in LocaRDS:
\begin{itemize}
\item \textbf{Transmitter location:} Location of the aircraft at the time of transmission. Locations are provided as WGS84 coordinates in decimal degrees. There are two altitude values provided, barometric and the geometric (WGS84). While geometric altitude is needed in most cases, in real-world air traffic control scenarios often only barometric altitude is known.

\item \textbf{Receiver location:} WGS84 coordinates of the receiving sensor. Note that locations can be inaccurate for different reasons, sometimes with high offsets to their real location \cite{schafer2018opensky}. However, it can be assumed that the vast majority of locations are accurate. In addition, we provide indicators marking verified locations, which we discuss further below.

\item \textbf{Sensor timestamp:} Timestamp measured by the sensor at the time of signal arrival in nanoseconds since beginning of the recording. Depending on the sensor type and the setup, this timestamp might be subject to drift of varying degree or even broken (see \cite{schafer2018opensky} for more information).

\item \textbf{Received signal strength indicator (RSSI):} Signal strength as measured by the receiver in dB with unknown reference. RSSI calibration depends on unknown factors such as antenna gain, device type, gain settings, etc. Since calibrating the RSSI is beyond the scope of this work, users who require calibrated values need to devise and apply a calibration method (e.g., based on the well-defined transmission power of ADS-B and the free-space path loss model).

\item \textbf{Server timestamp:} Timestamp measured by the server when the ADS-B position report was first seen, in microseconds since the beginning of the recording. This timestamp can be used, for instance, in time synchronization methods. Note, that it includes the unknown and highly varying internet delay between the sensor and OpenSky's server.

\item \textbf{Aircraft location accuracy indicator:} Binary per-aircraft indicator for the quality of the provided locations. The accuracy of locations provided via ADS-B depends on the availability of GPS on-board the aircraft. As ADS-B is still in its deployment phase, some aircraft report locations with an offset to their real location or with a delay.\footnote{Note that inaccurate aircraft locations may pose an interesting real-world use case for location verification algorithms.} This Boolean quality indicator helps users select reliable transmitter locations, e.g., as ground truth. If this indicator is false, then the location provided by the aircraft is known to have a low quality. A missing indicator means that the aircraft could not be verified due to bad geometry or a lack of reliable measurements.

\item \textbf{Sensor location accuracy \& synchronization indicator:} Binary per-sensor indicator for the time synchronization status and accuracy of the location information of a sensor. As timestamps and sensor locations are jointly used to assess their quality, errors cannot be attributed to either of them reliably. Hence, this flag is only true if the receiver-provided timestamps did not drift over the course of one hour and the sensor location could be verified based on these timestamps.

\end{itemize}

\begin{figure}[t]
    \centering
    \includegraphics[width=\linewidth]{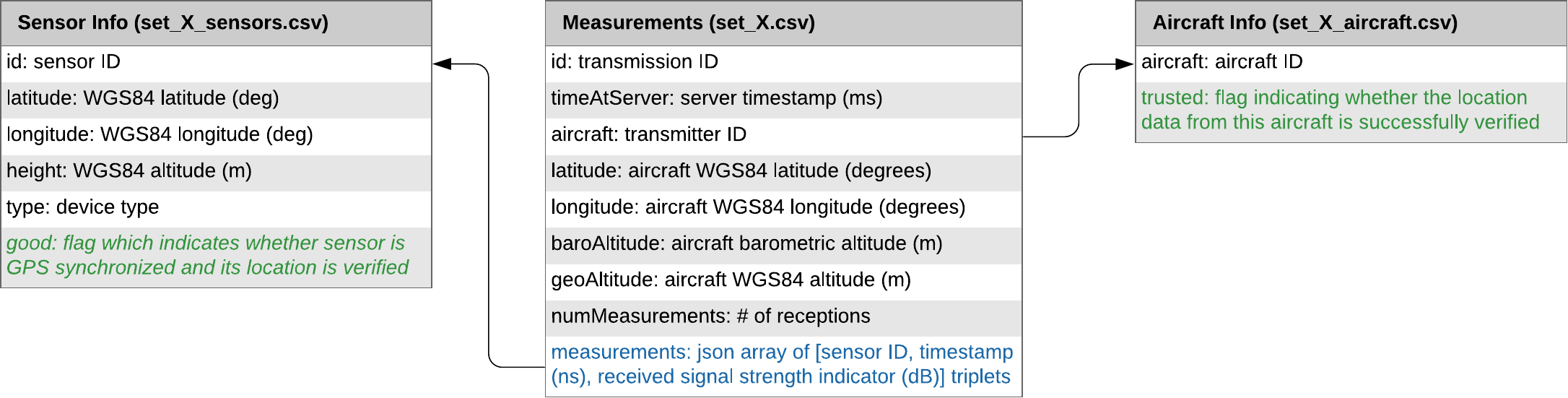}
    \caption{Structure of each 1h subset of LocaRDS. Each subset consists of three files. The main file \texttt{set\_X.csv} (X = 1...8) contains all measurements data. Each measurement is associated with multiple entries in the sensor information file (\texttt{set\_X\_sensors.csv}) and exactly one aircraft data validation result (\texttt{set\_X\_aircraft.csv}).}
    \label{fig:subset_structure}
\end{figure}

\subsection{File Structure and Format}

LocaRDS is split into eight subsets, each containing one hour of data (different weekdays / time of day). The duration was chosen as a trade-off between data volume per subset and having continuous flight movements covering longer distances of up to about 800~km. As a general file format, comma-separated values (CSV) was chosen for its wide support among data science tools and programming languages. Each subset with identifier X consists of three CSV files. The file \textbf{set\_X.csv} contains all ADS-B position reports that were received by at least two sensors, including the measurement data (timestamp and signal strength indicator) for each transmission. Locations and GPS synchronization status and location accuracy indicators of all sensors can be found in \textbf{set\_X\_sensors.csv}. Finally, \textbf{set\_X\_aircraft.csv} contains the location verification results indicating whether the location information provided by this aircraft has a high accuracy or not.

An overview of structure, formats, columns in each CSV file, and references between the files is given in Fig. \ref{fig:subset_structure}.

\subsection{Dataset Characteristics}

\begin{table}[tb]
    \caption{Information about the LocaRDS dataset. GPS/ADS-B sensor numbers are not additive as they contribute to several subsets.
    }
    \label{tab:subset_stats}
\begin{tabular}{@{}lllllll@{}}
\toprule
Set   & data points & deduplicated & verified   & sensors & GPS & GiB                               \\ \midrule
1        & 28,234,130  & 6,457,542    & 1,839,760  & 318     & 45          & 1.18 \\
2        & 28,717,685  & 6,535,444    & 1,680,956  & 317     & 45          & 1.20\\
3        & 28,749,671  & 6,569,830    & 1,996,987  & 318     & 44          & 1.20\\
4        & 28,215,712  & 6,348,679    & 1,810,382  & 317     & 43          & 1.17\\
5        & 26,313,445  & 6,111,569    & 1,452,447  & 314     & 41          & 1.11\\
6        & 27,360,671  & 6,309,260    & 540,953    & 313     & 40          & 1.15\\
7        & 27,514,781  & 6,345,589    & 779,524    & 313     & 39          & 1.16\\
8        & 27,395,507  & 6,187,378    & 1,793,618  & 309     & 42          & 1.14\\ \midrule
All & 222,501,602 & 50,865,291   & 11,894,627 & 323     & 46          & 9.31 \\ \bottomrule
\end{tabular}
\
\end{table}

In total, LocaRDS contains 222,501,602 single measurements from 323 sensors, split into eight 1h chunks. The total size of the CSV files is 9.31~GiB. As mentioned in Sec. \ref{sec:preparation}, all time and signal strength measurements referring to the same transmission are grouped during the deduplication process, resulting in 50,865,291 groups. In addition, whenever possible, the locations reported via ADS-B are verified based on the timestamps provided by GPS synchronized sensors. Overall, there are 11,894,627 transmissions from verified aircraft positions. An overview over the size and distribution of the data across the chunks is provided in Table \ref{tab:subset_stats}.

\subsubsection{Geographic Distribution}

All aircraft and sensor locations in LocaRDS are shown in Fig. \ref{fig:locards_locations}. The major European hubs such as London Heathrow, Frankfurt International, or Paris-Charles De Gaulle are clearly visible since most of the trajectories are converging towards them. The sensor density in LocaRDS is higher around major airports. This is highly beneficial for localization research since due to the line-of-sight limitation of 1090~MHz communications and the earth's curvature, range at lower altitudes is shorter. Hence, a higher sensor density is required close to airports where arriving and departing traffic is moving at low altitudes. Since these trajectories are more dynamic than in enroute traffic (which is mostly flying on straight lines), this low-altitude traffic might be of special interest to some researchers.

\subsubsection{Measurement Redundancy}

\begin{figure}[t]
    \centering
    \includegraphics[width=0.9\linewidth]{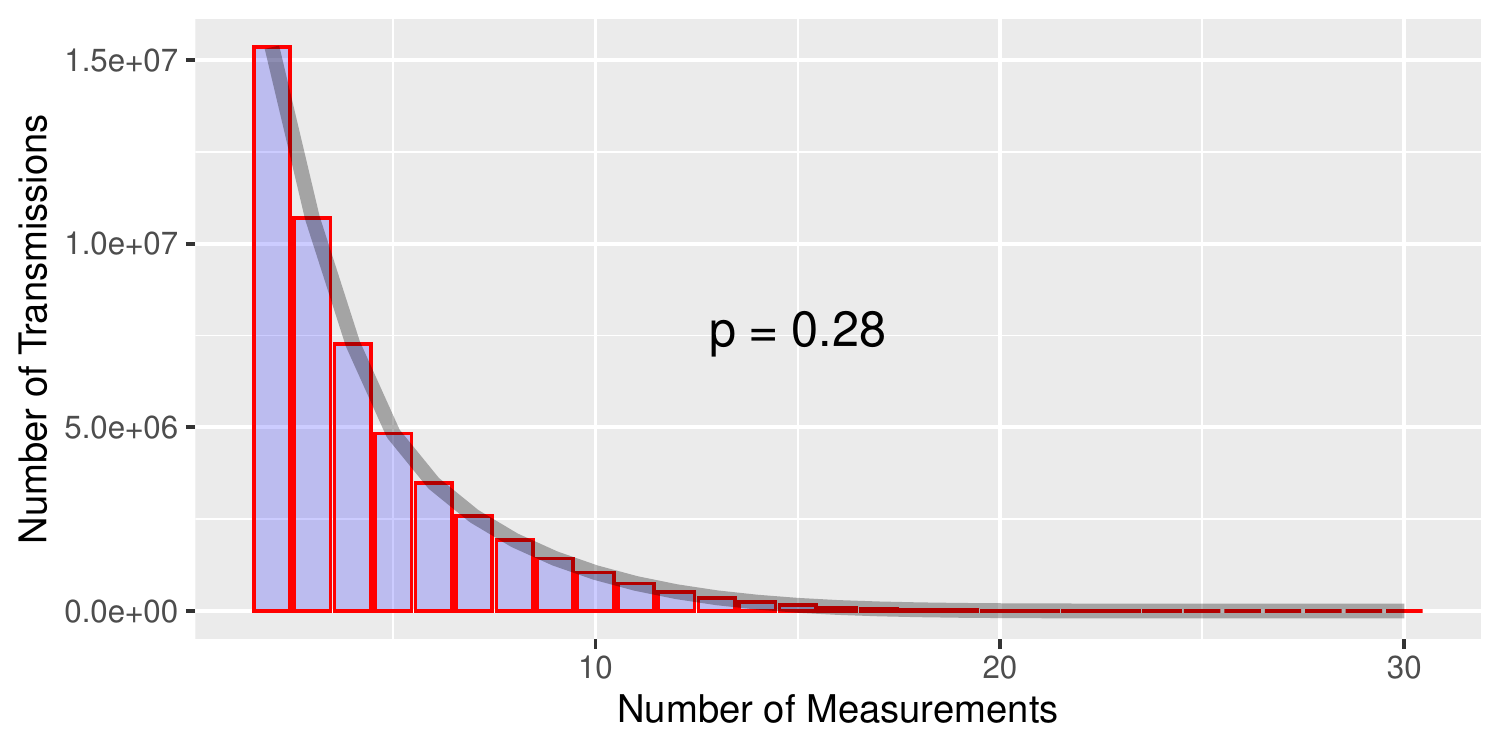}
    \caption{Distribution of number of measurements per transmission, following a geometric distribution with a success probability of 28\%.}
    \label{fig:redundancy_dist}
\end{figure}

\begin{figure}[t]
    \centering
    \includegraphics[width=0.9\linewidth]{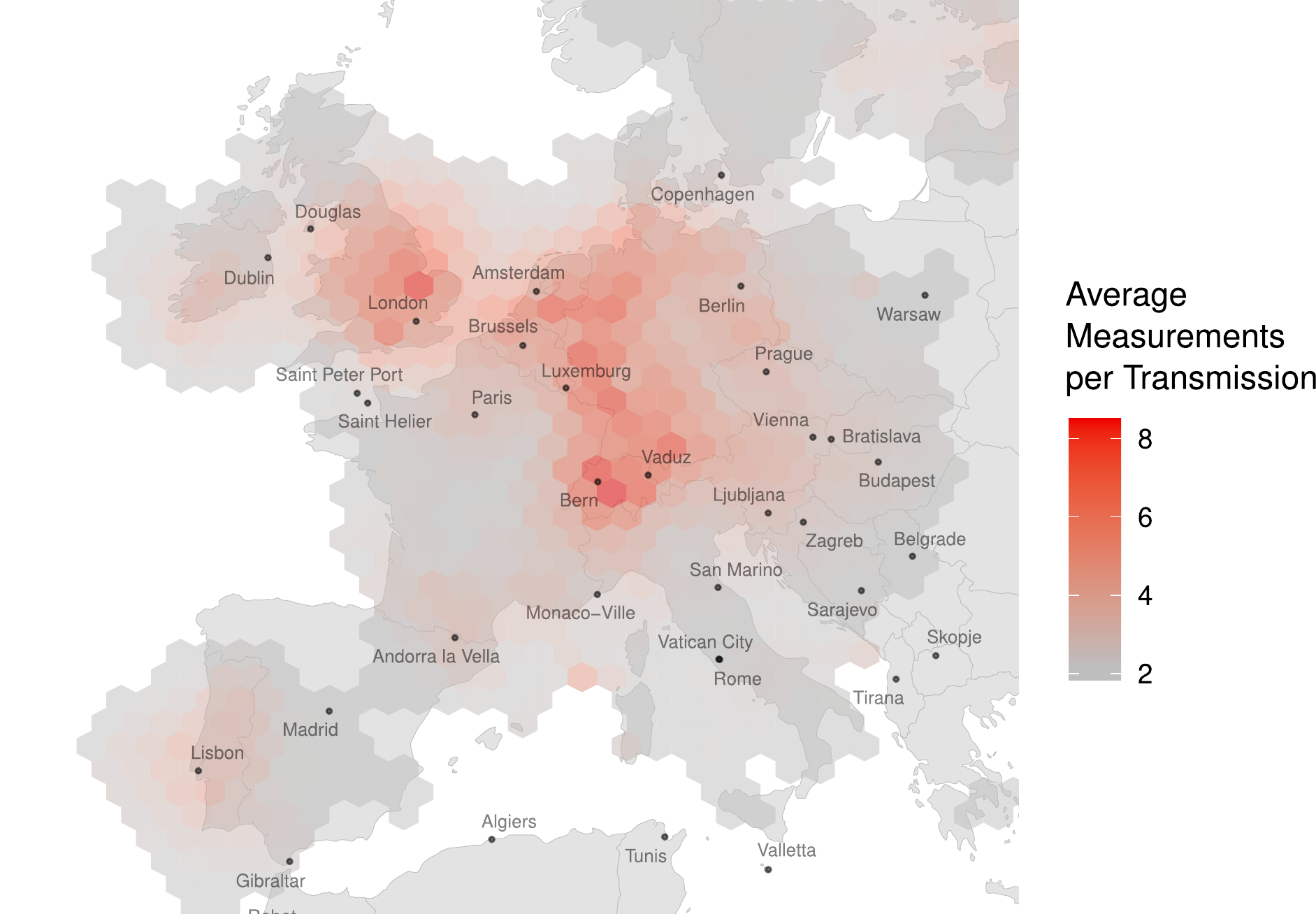}
    \caption{Distribution of the average number of measurements per LocaRDS transmission. The number of receivers for a single transmission ranges from 2 to 30, depending on OpenSky's coverage redundancy. It is highest in Central Europe with an average of 8.}
    \label{fig:redundancy}
\end{figure}

Each ADS-B position report included in LocaRDS was received by at least two sensors. As shown in Fig. \ref{fig:redundancy_dist}, the number of measurements per transmission strictly follows a geometric distribution with a maximum of 30 sensors receiving the same transmission. The geographic distribution of the coverage redundancy is shown in Fig. \ref{fig:redundancy}.

\subsubsection{Verified Data}

\begin{figure}[t]
    \centering
    \includegraphics[width=0.8\linewidth]{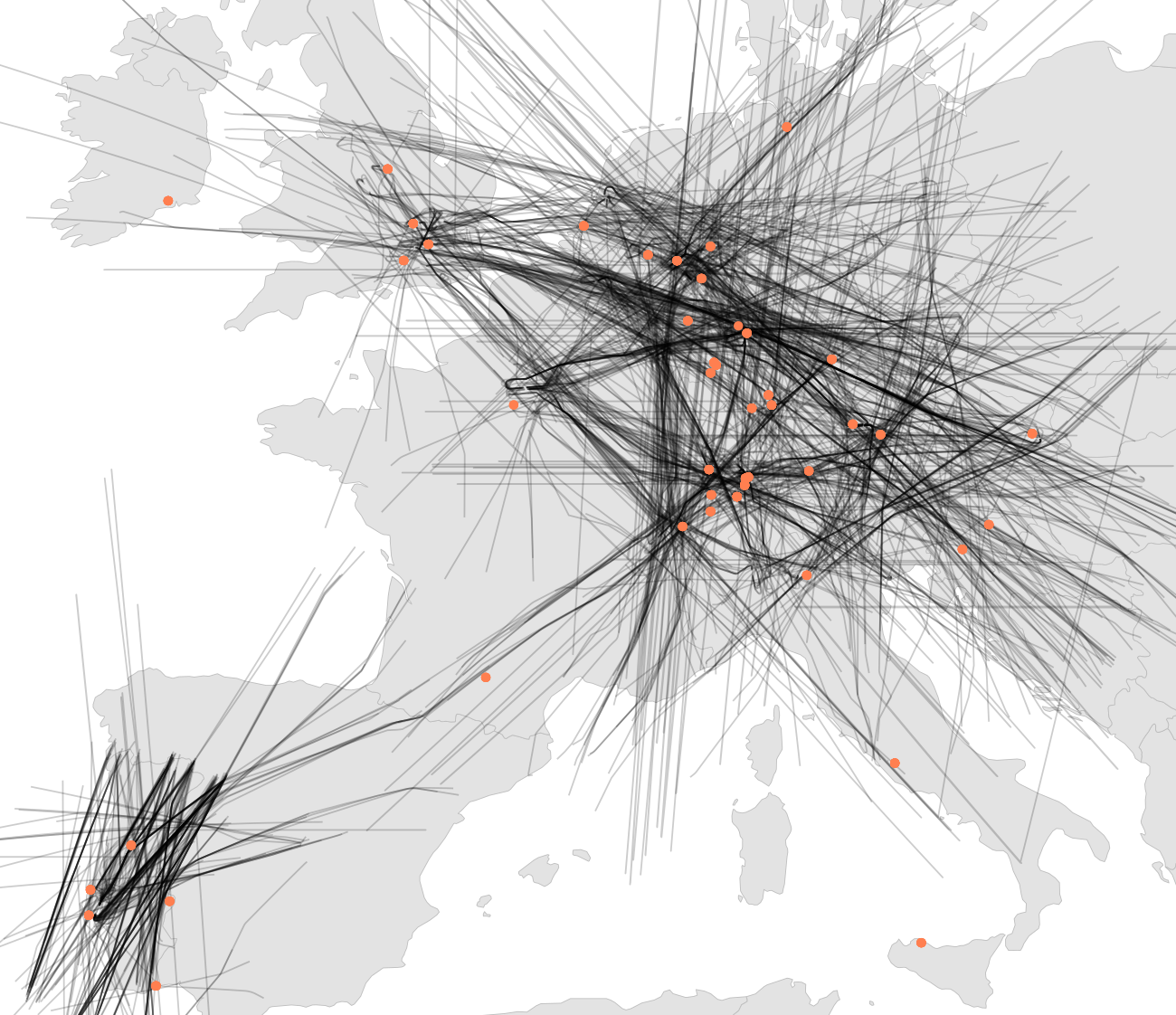}
    \caption{Verified aircraft and sensor positions in LocaRDS. The subset was created using the provided indicator that measuring the reliability of location and timing information from a specific aircraft and sensor.}
    \label{fig:locards_verification}
\end{figure}

About 12\% of the aircraft locations and 22\% of all measurements in LocaRDS are verified. However, this does not mean that 88\% of the reported positions are incorrect. In fact, a large fraction of the data (61\%) could not be verified due to bad geometric conditions (bad delusion of precision) or the lack of synchronized sensors. Since the data verification algorithm applied to LocaRDS (see Section \ref{sec:preparation}) relies on coverage redundancy and cross-checks of data coming from different receivers, the geographic distribution of the LocaRDS subset of verified aircraft positions and sensors is also concentrated around the hot spots in Fig. \ref{fig:redundancy}. The set of all verified aircraft and sensor positions is shown in Fig. \ref{fig:locards_verification}.

\subsubsection{Pseudonymized Data}

All contributors to OpenSky have the option to decide whether their accurate sensor location data may be distributed for research use or not. For those opting to participate, we have still pseudonymized their sensor IDs for LocaRDS. Likewise, we have pseudonymized all unique transponder IDs of the transmitting aircraft, although the distribution of historical track data are not generally a concern for most, as their wide availability shows. It has to be noted that some of this information could be recovered, if an adversary had access to the underlying OpenSky data or other large-scale databases and additional knowledge for example of the time frame of the data collection, but we consider the practical risks of this issue as negligible.

\section{Application Scenario: The Aircraft Localization Problem (ALP)}\label{sec:ALP}
To gauge the effectiveness of LocaRDS, we propose to evaluate solutions to the well-known outdoor positioning problem of locating aircraft. We formulate this as the problem to find the three dimensional position $ \mathbf{p} = [x, y, z]^T$ of an aircraft based on the signal characteristics of the communication. 

This type of localization or positioning can in principle be conducted with any communications signal sent out by an aircraft, including analog signals such as VHF radio\footnote{Although obtaining accurate timestamps is complex in the analog case.} or data link communication (e.g., the controller-pilot data link communications protocol, CPDLC). Without loss of generality, LocaRDS uses the widely supported ADS-B technology, which is mandated in most developed airspaces from 2020 and forms the heart of the next generation of air traffic control. ADS-B fulfills the requirements discussed in Section \ref{sec:requirements} and, in contrast to other technologies, is readily collected by many web trackers including OpenSky or Flightradar24. Thus, it offers not only sufficient data but based on its popularity also many target users that would benefit from improved ALP solutions, in particular in a crowdsourced setting. 

\subsection{Existing Solutions for the ALP}

We discuss the two fundamental physical layer characteristics used in the literature for solving the ALP, TDoA and RSS, in more detail. Both can be applied using LocaRDS.

\subsubsection{Time Difference of Arrival}

The most popular approach to aircraft localization is to use the time differences of arrival concept, where $n > 1$ ground sensors receive and match the same signal sent by an aircraft. At reception, every receiver timestamps the signal. The ToA measurements are then joined and the differences of all arrival timestamps $t_1,...,t_n$ between all $n$ involved receivers calculated. This is done for example by subtracting the earliest timestamp $t_{min}$ or using a fixed receiver of the set as anchor. This data then forms the basis for the TDoA approach, which as a surveillance technique is best known as \textit{multilateration} (MLAT).

Calculating the position from this TDoA data is done traditionally through solving a geometric problem, i.e. finding the intersection of the resulting system of hyperboloids. Solutions have been proposed using open-form, iterative and closed-form, direct algorithms, a short classification of these methods can be found in \cite{SIVP}. While the former require an initial estimation of the wanted position as input, the latter do not.

MLAT is a proven and well-understood concept used in civil and military surveillance. It serves as an operational method for ATC around airports and even smaller countries (e.g., Austria or Czech Republic). Academic works and aviation regulatory bodies have argued for MLAT being an ideal backup for primary radar systems, which are slowly being phased out due to cost, accuracy and reliability issues \cite{WAM}.

However, classic MLAT solutions suffer from a list of drawbacks, most notably expensive hardware to enable highly accurate timestamps and tight synchronization. Both are a strict necessity for MLAT algorithms as they are highly sensitive to noise, in particular in uncontrolled receiver placements where the geometric characteristics are not optimized  \cite{strohmeier2018k}.

\subsubsection{Received Signal Strength}

The alternative to TDoA-based localization is to directly determine the distance between a target and multiple reference locations. The target location is then estimated by finding the intersection of the resulting circles (instead of hyperboloids) around the reference locations. 

The distances can be obtained by measuring a signal's time of flight or its strength. The former can be obtained through the round trip time (as in classical radar) or by relying on a tight clock synchronisation between the target and the localising infrastructure. Both methods are not available through ADS-B and generally more limited in their applicability since they either require active communication or expensive synchronisation. The RSS, on the other hand, offers a cheap alternative as it is measured by most available receivers.

By measuring the strength of an incoming signal and by knowing or estimating the transmit power of the aircraft, the distance between sender and receiver can be estimated based on their difference, i.e., the \emph{path loss}. One notable drawback of the RSS is, however, that its accuracy depends on many potentially unknown factors, the radio environment, and (analogous to TDoA) the measurement resolution \cite{Whitehouse07}.

Besides direct ranging measurements, RSS-based localization approaches often use indoor radio maps. This is intuitively more difficult to recreate with fast-moving aircraft spread out over long distances in highly-dynamic environments (due to weather, buildings and other influences). Building radio maps further requires a setup phase and separate infrastructure, which cannot be offered through a reference dataset.

\subsection{LocaRDS Reference Localization Implementation}

As it is difficult to comprehensively provide tight time synchronization in crowdsourced networks with low-cost receivers, our reference (TDoA-based) localization implementation is composed of two parts: an opportunistic synchronization algorithm and a hyperbolic  localization algorithm. 

First, the synchronization mechanism exploits ADS-B position messages to estimate the time offset between any two of sensors which share air traffic communication. After that, the \textit{paired sensors} can be used to estimate the positions of aircraft emitting any type of signal with measurable ToA.

This combined algorithm is capable of locating ADS-B transmitters or, in general, aircraft emitting any kind of radio signal for which is possible to measure the time of arrival. Moreover, it has been specifically designed for crowdsourced sensor networks as it does not need an external synchronization method (e.g., GNSS, reference transponder, NTP or others) and is robust to outliers and anomalies (such as offline sensors or measurement integrity issues).

\subsubsection{Sensor Pairing and Offset estimation}
It is hard to synchronize all sensors in a controlled wide-area or global sensor network to a sub-microseconds level; it becomes practically impossible in case of crowdsourced networks with no control over heterogeneous hardware and sensor setups. Additionally, each sensor can be switched on and off at any moment.

For these reasons, we exploit an empirical method to synchronize the data received from the network. For each pair of sensors receiving the same position message from an aircraft, the clock offset is estimated measuring the ToAs of the message at each sensor ($t_i$,$t_j$), knowing some ground truth about the aircraft position $\mathbf{p}$ (e.g., from previous ADS-B messages) and the positions of the sensors ($\mathbf{s_i}$,$\mathbf{s_j}$):

\begin{equation}
\Delta t_{i,j}=  \left(\frac{\left\|\mathbf{s}_i-\mathbf{p}\right\|}{c}-\frac{\left\|\mathbf{s}_j-\mathbf{p}\right\|}{c}\right)- (t_{i}-t_j)
\label{tdoa}
\end{equation}

By repeating this measurement for all messages from aircraft in common view, a time series $ \lbrace \Delta t_{i,j}\rbrace$ for the paired sensors' offset is obtained and tracked to improve the offset estimation accuracy and to forecast its value for any point in time.\footnote{Any common tracking algorithm such as \cite{brookner} can be used, a possible approach for this specific application is given in \cite{Leonardi2019}.} If more than two stations are in common view of the same aircraft, the offsets for any two pairs can be estimated.

It is noted that with this method, the network does not have full synchronization, i.e., there is no single common reference time. Moreover, sensors that do not share any ADS-B traffic cannot be  synchronized, which is not a problem for our application as these are also not helpful in positioning the aircraft. On the other hand, this opportunistic method is simple and robust as it requires neither a complex time transfer mechanism, a common shared reference time system, nor physical reference stations.

In crowdsourced sensor networks, this provides several concrete advantages. First of all, each station can have a service interruption, transmitting no or incorrect data,  without affecting other pairs' synchronization. Second, even in cases where an aircraft transmits wrong data, it will affect only the sensors in its range. In this case, if the stations have also other aircraft in their common range, the incorrect data will be discarded as outliers through a subsequent pass of the tracker using an innovation test (see \cite{Leonardi2019}, \cite{galleani_tavella}). When the innovation (i.e., the difference between expected measurement and incoming measurement) is larger than a given threshold, the incoming measurement is not used and the offset is extrapolated by the previous state. If the threshold is exceeded several consecutive times, the tracking algorithm is re-initialized.

\subsubsection{Aircraft Hyperbolic Localization using Paired Sensors}
\label{sec:localization_algorithm}

Classical MLAT algorithms assume that all sensor clocks are synchronized to the same reference station, providing full network synchronization. Under this assumption, four ToA measurements (i.e., three synchronized TDoA measurements) are sufficient for a 3D-localization of the target. As discussed this assumption does not hold in a crowdsourced network and our proposed approach only synchronizes sensor pairs. 

In this case, a  modified hyperbolic localization algorithm can be used. For any aircraft to be localized, all possible paired sensors in view are identified, rearranging Eq. \ref{tdoa} as follows:

\begin{equation}
TDoA_{i,j}=(t_{i}-t_j)=  \left(\frac{\left\|\mathbf{s}_i-\mathbf{p}\right\|}{c}-\frac{\left\|\mathbf{s}_j-\mathbf{p}\right\|}{c}\right)- \Delta t_{i,j}  \hspace{1cm}
\label{tdoa2}
\end{equation}

Now $\Delta t_{i,j}$ is estimated by the tracking filter and $\mathbf{p}$ is the unknown to be estimated. If more than three pairs of synchronized stations are in range of the aircraft, the corresponding equations can be used to find the airplane position. 

\begin{figure}[t]
	\centering\includegraphics[keepaspectratio=true,width=0.9\columnwidth]{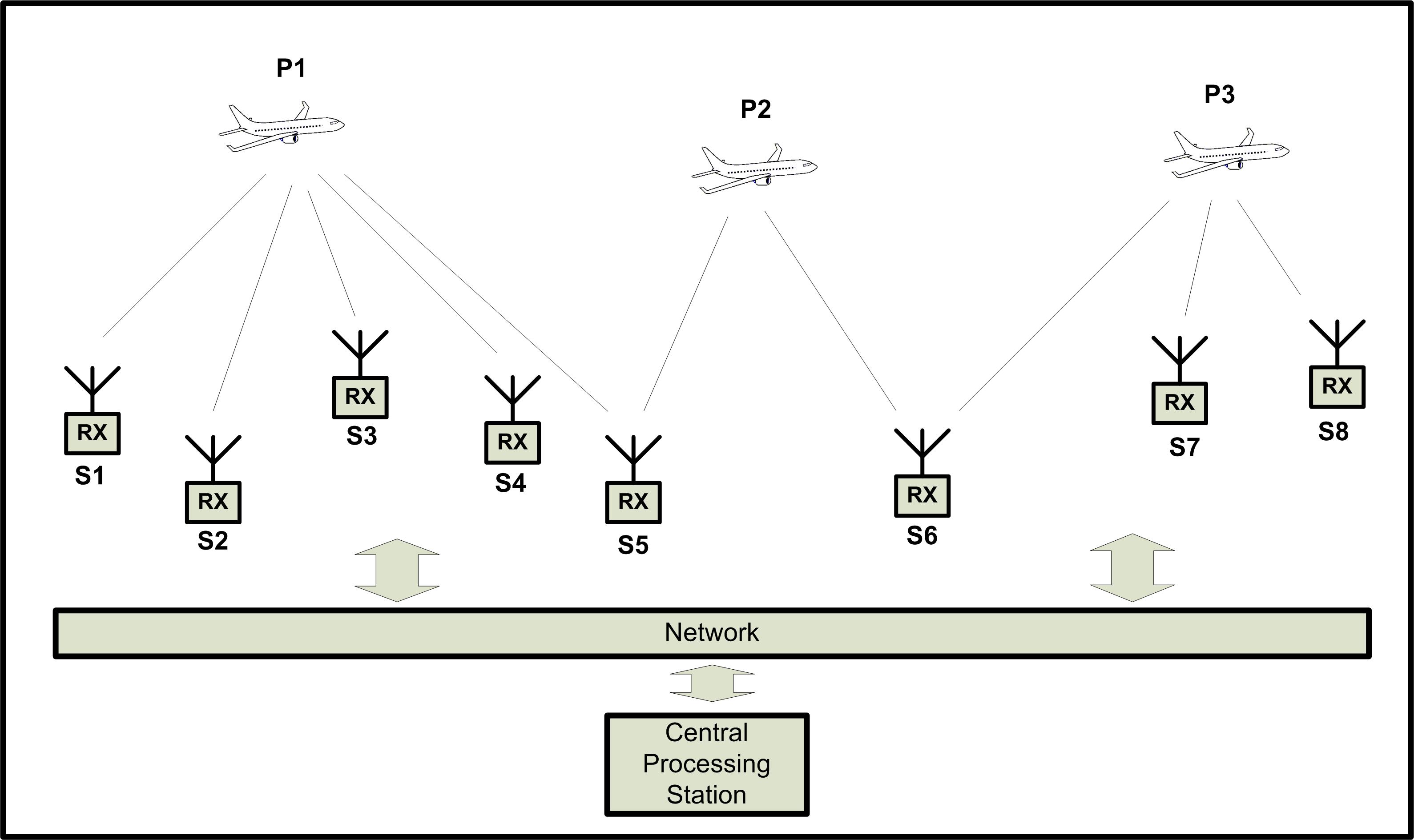}
	
	\caption{Illustration of reference implementation with sensor pairing.}
	\label{syst}
\end{figure}

This process is exemplified in Fig.~\ref{syst}. The aircraft in position $P3$ can be used to synchronize the pairs $\left\lbrace S6,S7\right\rbrace$, $\left\lbrace S6,S8\right\rbrace$ and $\left\lbrace S7,S8\right\rbrace$ and the aircraft in $P2$ to synchronize the pair $\left\lbrace S5,S6\right\rbrace$; the four pairs $\left\lbrace S6,S7\right\rbrace, \left\lbrace S6,S8\right\rbrace, \left\lbrace S7,S8\right\rbrace,\left\lbrace S5,S6\right\rbrace$ can then be used to localize any airplane whose communication is received by stations ($S5,S6,S7,S8$), producing four different equations. 

While some equations could be linearly dependent (for example, $\left\lbrace S7,S8\right\rbrace$ is a linear combination of  $\left\lbrace S6,S7\right\rbrace$, $\left\lbrace S6,S8\right\rbrace$), if at least three equations are independent of each other, the aircraft position can be estimated. 

To solve this system of non-linear equations, any classical method can be used. For example, if an approximated starting estimate of the aircraft's position is known, the Newton-Raphson method can be applied. 

In our reference localization implementation, a simple method to fix the initial estimate is used:
\begin{enumerate}

	\item If no estimation of the target is present from previous observations, the average of the sensor positions in range of the aircraft is used. This average can be weighted by  signal strength (cf. the Centroid algorithm \cite{SHAKSHUKI2019501}).
	
	\item Otherwise, the last estimate of the target position is used.
\end{enumerate}

The performance of our reference implementation of the ALP is evaluated as part of the case study in the next section.

\section{Evaluation}\label{sec:evaluation}
We now propose some metrics that can be used to evaluate algorithms using LocaRDS. There is no claim to completeness, on the contrary we expect this do be an ongoing process, which we will actively support on through the OpenSky website. The metrics chosen for the scientific evaluation of the ALP should be as broadly applicably to the different scenarios and approaches as possible. 

\subsubsection{Localization Accuracy}
The key metric for localization is the accuracy with which the position of the target is predicted. While the utility of aircraft localization depends very much on the context and the use case,\footnote{There are 12 categories of navigational accuracy defined by the International Civil Aviation Organisation, from $< 3$ meters to $> 10$ nautical miles.} higher accuracy is strictly better. We chose the root mean square error (RMSE) between the real aircraft position as reported by the ADS-B ground truth and the contestants' predictions for our main ranking metric. The RMSE has been widely included as a standard metric to compare the predictive performance of different localization models (see, e.g., \cite{lymberopoulos2015realistic}). 

\subsubsection{Dataset Coverage}

The second consideration concerns the coverage of the evaluation datasets, i.e., how many of the data points of LocaRDS were chosen to be predicted. While ideally all samples would have a prediction, this is not practical for several reasons. For example, some methods may need initial samples to calibrate and also regularly re-calibrate. Furthermore, there is also value in correctly choosing to not predict bad or uncertain samples in order to minimize outliers and improve the average localization performance.  However, it is obvious that with equal localization accuracy higher coverage is strictly better.

Concretely, we required a minimum sample coverage of 50\%, which should on average satisfy any non-tactical applications of the ALP, i.e. those where update rates of aircraft positional information of more than 1 second are allowed. However, other values can sensibly be chosen based on the application requirements and also depending on the sensor coverage in a given geographical region. 

\subsubsection{Further Considerations}
Due to the variation in the distribution of uncertainty and quality of measurements in LocaRDS, it is clear that there can be trade-offs between choosing a high coverage and a high accuracy. Besides requiring a minimum coverage, this trade off can also be quantified for a provided solution through applying a penalty directly towards the accuracy scoring. By assuming a fixed high localization error for any missing observation, the RMSE is increased, incentivising the contestants to provide a higher number of observations. However, the effectiveness of the penalty is highly dependent of the quality of the provided solution --- if the penalty is set below the RMSE, it will actually improve the quality score and thus set a false incentive to leave out observations. Hence, we adjusted this penalty based on the average quality of the solutions and our reference implementation. The output is a \textit{truncated root mean square error}, which we used to conclusively rank the participants.

A second consideration is centered around the runtimes of possible solutions. While the speed of localization algorithms is not crucial in many application scenarios, and most modern computing architectures should be able to fulfil any real-time constraints, it may still be insightful to analyze. Variations in training times for ML-based solutions may impact the choice of algorithms in situations where regular re-training is required. Similarly, lightweight algorithms for distributed resource-constraint edge computing are a relevant application for example for crowdsourced flight tracking networks. 
\subsubsection{Results}

In comparison, our reference implementation based on traditional multilateration shows that good aircraft localization results can be achieved with crowdsourced measurements based on cheap off-the-shelf hardware. It targeted measurements with at least 3 receivers, as is geometrically required for pure MLAT, and thus achieved a coverage of 45\% and a truncated RMSE of 682.38 meters.

\section{Conclusion}\label{sec:conclusion}
In this paper we introduced LocaRDS, a reference dataset for localization and positioning research. We derived the requirements for such a reference dataset and showed how LocaRDS can successfully be used to test, analyze and directly compare different localization techniques. While in the present work we applied this knowledge to the open research problem of the aircraft localization problem in crowdsourced networks, we postulate that its appeal is much broader. Thus, for future work, we hope that many researchers consider LocaRDS not only for problem of aircraft localization but for their own problems in the fields of positioning, self-positioning and location verification. We believe that transparent competition can offer these fields a better way forward and results obtained with LocaRDS a broader scientific and practical application.

\balance
\bibliographystyle{IEEEtran}
\bibliography{bibliography}

\begin{IEEEbiographynophoto}
{Matthias Schäfer} is a lecturer and researcher at the distributed computer systems lab (DISCO) at the University of Kaiserslautern, Germany. Until 2018, he was a Ph.D. student supervised by Prof. Dr.-Ing. Jens B. Schmitt. Before that, he worked at armasuisse S+T and visited the University of Oxford in 2012. He is also a co-founder and board member of the OpenSky Network association and managing director of SeRo Systems GmbH.
 \end{IEEEbiographynophoto}
 
 \begin{IEEEbiographynophoto}
{Martin Strohmeier} is a Junior Research Fellow at the University of Oxford and a Scientific Project Manager at the Cyber-Defence Campus in Switzerland. Before coming to Oxford for his PhD, he received his MSc from TU Kaiserslautern, Germany and worked as a researcher at Lancaster University's InfoLab21 and Lufthansa. He is also a co-founder and board member of the OpenSky Network association.
 \end{IEEEbiographynophoto}
 
\begin{IEEEbiographynophoto}
{Mauro Leonardi} is currently an Assistant Professor with the University of Rome Tor Vergata, Rome, Italy. He teaches radiolocation and navigation systems and fundamentals of radar. His research interests include radar, air traffic control, satellite navigation, integrity and signal analysis, new surveillance systems for ATC, positioning, and localization algorithms.
 \end{IEEEbiographynophoto}

\begin{IEEEbiographynophoto}
{Vincent Lenders} is head of the Cyber-Defence Campus at armasuisse, Switzerland. He received the MSc and PhD degrees in Electrical Engineering and Information Technology from ETH Zurich. He was a postdoctoral research fellow at Princeton University. Dr. Lenders is a co-founder and on the board of the OpenSky Network and Electrosense associations.
 \end{IEEEbiographynophoto}

\end{document}